\documentstyle[multicol,aps,prb]{revtex}
\tightenlines

\renewcommand{\appendix}{%
 \setcounter{section}{0}%
 \setcounter{equation}{0}%
 \renewcommand{\thesection}{{APPENDIX} \Alph{section}}
 \renewcommand{\theequation}{\Alph{section}.\arabic{equation}}%
}

\begin{document}
\draft
\preprint{LA-UR-97-5086}

\title{\bf Onset of polymer entanglement}
\author{Shirish M. Chitanvis}
\address{
Theoretical Division, 
Los Alamos National Laboratory\\
Los Alamos, New Mexico\ \ 87545\\}

\date{\today}
\maketitle
\begin{abstract}
We have developed a theory of polymer entanglement
using an extended Cahn-Hilliard functional, with two extra terms.
One is a 
nonlocal attractive term, operating
over mesoscales, which is interpreted as giving rise to entanglement,
and the other a 
local repulsive term indicative of excluded volume
interactions. 
This functional can be derived using notions
from gauge theory.
We go beyond the Gaussian approximation, to the one-loop level, to show
that the system exhibits a crossover to a state of entanglement as
the average chain length between points of entanglement decreases.
This crossover is marked by {\it critical} slowing down, as the effective
diffusion constant goes to zero.
We have also computed the tensile modulus of the system, and we find a
corresponding
crossover to a regime of high modulus.
The single parameter in our theory is obtained by fitting to available 
experimental data on polystyrene melts of various chain
lengths.
Extrapolation of this fit yields a model for the cross-over to entanglement.
The need for additional experiments detailing the cross-over to the
entangled state is pointed out.
\end{abstract}


\begin {multicols}{2}
\section{Introduction}

While it has long been known that entanglement in homopolymers
has an important effect on its strength, a thoroughly satisfactory
theory of polymer entanglement is still a topic of current research.
The classic experimental work of Moore and Watson\cite{moore}
showed that the bulk modulus of cross-linked natural rubbers
depends inversely on the average chain length ($N_c$) between
cross-links in the 
system, and that end corrections
become negligible as the total average molecular weight $\mu$
gets very large.
They pointed out the analogy between chemical cross-linking and
physical entanglement.
Thus their work applies in a qualitative sense to entangled systems as well.
Their work extended the earlier pioneering work of Flory et al\cite{flory,doi}.

Edwards developed the tube theory of the effect of entanglement on 
elastic moduli of homopolymers using deGenne's idea of reptation\cite{doi}.
This theory
showed how entanglement enhances the tensile modulus of a homopolymer.
He also developed a more detailed model of entangled ring polymers 
using notions from knot theory\cite{doied}.
The basic idea behind this theory is an analogy between 
certain mathematical invariants describing intertwined
loops and magnetic fields induced in wires by current-carrying loops. 
Prager and Frisch\cite{prager} worked on this notion as well, as did 
Koniaris and Muthukumar\cite{koniaris}.

More recently, interest has turned towards computer
simulations of polymer networks, involving various levels of molecular
detail, to understand the effect of entanglement on the strength
of homopolymers.
As examples, we mention the work of Termonia\cite{termonia} 
and Bicerano\cite{bicerano}, who use
phenomenological models of polymer networks to study
their viscoelastic properties.
Comparison with experimental data shows a varying degree of success,
depending on the particular system studied.
Holtzl et al\cite{holtzl} use the 
more basic fluctuating bond theory to
model a network of polyethylene strands to show that 
entanglement leads to non-affine displacements under large tensile strains.

In an earlier paper,\cite{shirish} we developed a gauge theory 
of self-assembly, and utilized renormalization group ideas to
study the onset of self-assembly
in diblock copolymers.
In this paper, we shall pursue a similar continuum approach
to understand entanglement.

Intuitively, one can see that entanglement could be described by
assuming two
extra terms in the Cahn-Hilliard functional,\cite{cahn} one of which is
a nonlocal attractive term, which gives rise to entanglement
and the other, a soft-core
local repulsive term which arises from the fact that the strands
cannot cut across each other.
We connect the parameters which appear in our theory to
the underlying chain parameters with a simple model.
We have shown (see Appendix A) how such a functional can be
derived naturally using notions from gauge theory.

The results derived from our continuum formulation will be seen
to be reminiscent of the chain-theory approaches 
of Kavassalis and Noolandi\cite{kn} for
flexible polymer networks, and that of Kroy and Frey\cite{kf}
for semi-flexible networks.
They utilized a mean-field approach to locate the transition to
the state of entanglement.
Our theory is also somewhat similar to the paper of
Castillo and Geldart\cite{castillo}, who use a 
$\phi^3$ field theory, coupled to the
replica trick (in the mean field approximation) of Deam and Edwards\cite{ded} 
to study the vulcanization transition.

We shall utilize a field theoretic approach
and go beyond the Gaussian approximation
in this paper to show that the onset of the state of
entanglement is a crossover phenomenon, rather than a pure
phase transition, in that the effective diffusion
constant goes to zero at the transition point,
but the correlation length and the structure factor do not diverge.
A physical reason that fluctuations become important near the
onset of the state of entanglement is that the average chain
length between points of entanglement gets smaller, while in a
vulcanized polymer, the cross-links make the system increasingly stiff.
This underscores a difference between vulcanization
and entanglement-
An entangled network of polymers is more dynamic than a
vulcanized network.
The mean field approximation is expected
to be correct\cite{degennes} for the vulcanization transition.

We have also computed the tensile modulus of the system.
Corresponding to the critical slowing down discussed above, we
find a crossover in the modulus to a regime of high values.
Fits to available experimental data shows how the single parameter in our 
theory can be parameterized in terms of the molecular weight of the
system.
Extrapolation of the parameterization we have provided in this paper displays the
cross-over to the entangled state.
The need for further experiments detailing this cross-over is pointed out.


\section{A field theory of entanglement}

The continuum, mesoscale approach adopted in this paper 
assumes that we have performed some spatial averaging
of our polymeric system, so that the {\it order parameter}
is the local concentration of the polymers.
Our mesoscopic theory of entanglement in polymers is based on the
intuitive notion that physical entanglement 
can be captured by a non-local attraction between
the polymers which causes them to remain in proximity.
There must be a balancing, repulsive local energy term which  
says that the polymers cannot cut across each other.
The starting point of our mesoscale theory is
an internal energy functional which is quadratic in the gradient of
the local number concentration.
For the moment, we will
consider isolated systems, so that the 
quantity that is conserved is the internal energy\cite{callen}.
We will shortly consider entropy effects as well.
Consider the following form for the energy functional:

\begin{equation}
\beta U_0 =  \beta \int u_0 (c({\bf s})) {{\rm d}}^3 s 
\label{2.1}
\end{equation}
\begin{equation}
\beta = {{1}\over {k T}}
\label{beta}
\end{equation}

\begin{equation}
\beta u_0 ({c}({\bf s}))  =  \left({g\over2}\right) 
      {{\partial {c}({\bf s})}\over{\partial s_i}}
                          {{\partial {c}({\bf s})}\over{\partial s_i}}
\label{2.2}
\end{equation}

where 
repeated indices are summed over, 
${\bf s}$ is a dimensionless co-ordinate variable,
$k$ is Boltzmann's constant, $T$ is the temperature, and $c$ is the number
concentration of the specie.
The local concentration $c$ is normalized by dividing by 
some characteristic inverse volume.
The constant $g$ is analogous to a dimensionless diffusion constant.
Such energy functionals have been considered over many years as
contributing to the total internal energy of both unary and binary mixtures.
\cite{cahn}
We will use this form as our starting point to suggest a more
complete energy functional:

\begin{eqnarray}
\beta U_{{\rm eff}} &&= \beta U_0
  +\left({\alpha^2\over 2}\right) \int {\rm d}^3 s~{{c}}({\bf s})
                                            {{c}}({\bf s}) \nonumber\\
   &&- \left({\gamma \over {2 \pi}}\right) \int {\rm d}^3 s \int {\rm d}^3 s'~
                     {{c}}({\bf s}) {exp(-\delta
                                        \vert {\bf s} - {\bf s}' \vert)\over
                                              {\vert {\bf s} - {\bf s}' \vert}
                                           }
                                                 ~   {{c}}({\bf s}')
\label{gauss}
\end{eqnarray}

where $\alpha^2, \gamma, \delta$ are positive constants.
The local repulsive term is indicative of the fact that 
polymers cannot cut across each other.
This is in effect a soft-core repulsion term, and the softness arises because
we are studying a homopolymer network at a mesoscale,
where polymers may pass by each other, without actually cutting
across each other.
The nonlocal attractive term gives rise to entanglement, as it
causes portions of the network within the screening distance
$1/\delta$
to be attracted to each other.
Equation \ref{gauss} is the basic statement of our theory.
Note that the two terms we just discussed have signs opposite
those of corresponding terms in theories of
self-assembly\cite{shirish}.
The non-local attractive term invokes the notion that entanglement of
polymers must lead to knotty configurations\cite{stepanek}.
This attractive term can be seen intuitively to lead to the notion of
trapping, and as such is slightly similar to treatments of
entanglement in the literature which use Arrhenius-like rate theories
to provide for escape of polymers from local entangled arrangements.

In what follows,
we shall set $\gamma = \alpha^4$, and $\delta^2 = \sqrt 2 \alpha$,
with $\alpha^2 = g^2/2$.
A strong motivation for this choice of parameters
is provided in Appendix A, where we use notions from
gauge theory to derive Eqn.\ref{gauss}, with the parameters
having the forms given above.
Another explanation for such a choice is as follows.
With our choices for the parameters, U$_{\rm eff}$ in
momentum space may be written as:

\begin{equation}
U_{\rm eff} = \int {d^3k \over (2 \pi)^3} \hat c^*(k)
       [ -\sqrt 2 \alpha k^2 - \sqrt 2 \alpha k^2/(1 + \sqrt2 k^2/\alpha)]
             \hat c(k)
\label{Uk}
\end{equation}

where the carats indicate a Fourier transform.
Thus we see that the choices made for the parameters are equivalent to
generalizing the diffusion constant 
$g \equiv \sqrt 2 \alpha \to \sqrt 2 \alpha 
[1 + 1/(1 + \sqrt2 k^2/\alpha)]$, i.e., a
non-local diffusion constant is obtained.
If we now extremize the functional, the Euler-Lagrange equations
can be written in conservative form as:

\begin{eqnarray}
\vec \nabla \cdot \vec I(\vec s) &&= 0 \nonumber\\
\vec I(\vec s) &&= \vec \nabla \int {d^3k\over(2 \pi)^3} 
                \exp(i \vec k \cdot \vec s) 
                   [1 + 1/(1 + \sqrt2 k^2/\alpha)]\hat c(k)
\label{numcon}
\end{eqnarray}

where $\vec I(\vec s)$ can be interpreted in the conventional manner
as a mass current.
The divergence-free nature of this current
makes it clear that with our choice of parameters, our internal
energy functional preserves number conservation.
This is quite appropriate, since the internal energy
is the quantity which is conserved for isolated systems.
For an arbitrary choice of parameters, the Euler-Lagrange
equations have the form:
$\vec \nabla \cdot \vec I'(\vec s) = {\rm Source/Sink~Terms}$,
indicating that number conservation can be a problem.

While our choice of parameters may appear to be overly restrictive,
it turns out to be sufficiently rich 
to provide a description of the onset of entanglement in polymers.
We will not explore more general sets of parameters in this paper. 

Before we can compare our theory with experimental data, we
need to consider the fact that our system is not really
isolated, and may be in contact with an energy reservoir, 
perhaps as it is being acted on by mechanical forces
in a stress experiment.
For a system in contact with an energy reservoir, the 
quantity that is conserved is the Helmholtz free energy\cite{callen}
$A = U_{{\rm eff}} - ST$, where $S$ is the entropy of the system.
The entropy of our system will be approximated in the conventional
manner\cite{raveche}: 

\begin{equation}
-{S\over k} = \int {\rm d}^3s~ c({\bf s})~{\ln}[c({\bf s})] 
\label{s}
\end{equation}

Note that we have ignored a term in the above expression which is
linear in the normalized concentration field.
This term can be absorbed into the definition of the usual Lagrange
multiplier constraint for number conservation.
This additional constraint is necessary, over and above the
considerations which led to Eqn.\ref{Uk}, because we are now
considering the Helmholtz free energy rather than just the internal
energy. 
In the homogeneous mean field approximation, the chemical potential
can be easily shown to be zero.
We shall utilize this approximation to facilitate computations.
This entropy term provides the free energy a single minimum.
To further ease computations, we shall expand $c \ln(c)$ in a power
series 
about the characteristic inverse volume 
$\l^{-3} (= 1)$ in our dimensionless units$)$, 
retaining terms up to fourth order:

\begin{equation}
(1+c) \ln (1+c) \approx 
        c 
     + {c^2\over 2 } - {c^3 \over 6 } +{c^4 \over 12}
\label{cubic}
\end{equation}

This expansion yields one minimum, just as the exact expression
for the entropy.
Consequently, we do not expect this system to display a phase transition,
but rather a crossover from an un-entangled state to a state of
entanglement.
Finally, we note that in our present theory, entropy yields the
crucial nonlinear terms which will describe the crossover to a state
of entanglement, in contrast to our gauge theory of
self-assembly\cite{shirish} 
where entropy did not play a dominant role.

We define the two-point Green's function as usual via
${\cal S}(\vec x,\vec x') = Lim_{J\to 0} (\delta^2/\delta J(\vec x)
\delta J(\vec x')) Q[J] $,
where 
$Q[J] = \int {\cal D}c \theta(1+c)
  \exp[-\beta (U_{{\rm eff}} - S T) -\int d^3 s J(\vec s) c(\vec s)]$,
where $\theta(1+c)$ is a step function that indicates a restriction to 
physically acceptable values of the concentration.
In practice, we shall be restricting our attention to small
deviations of $c$ from its average, so that the step function is
implicitly accounted for during calculations.
In the quadratic approximation, the structure factor is:

\begin{equation}
{\hat{\cal S}}_0(k) = {1\over {1 + \alpha' k^2 + \alpha'^2 k^2/(1+2 k^2/\alpha')}}
\label{sk}
\end{equation}

where $\alpha'=\sqrt 2 \alpha$.
Equation \ref{sk} displays a peak at the origin,
as one might expect from the fact that entanglement
creates blobs which are
distributed at random within the system.
The width of the peak indicates an inverse of the correlation length
between the blobs.
With this physical interpretation, $\sqrt \alpha$ is a measure of
the distance between concentration fluctuations (i.e. between 
points of entanglement).
The decay of ${\hat{\cal S}}_0(k)$ is affected by the value of $\alpha$.
As $\alpha$ decreases, the structure factor looks more diffuse.
Thus,
a decrease in $\alpha$ signifies a shift to a state of higher entanglement,
as the concentration of entanglement points increases.

Our results can be understood compactly in terms of the
parameter $\alpha$, or equivalently, $g$, which may be identified
withe the self-diffusion coefficient of a polymer.
As dictated by the discussion in section IV later in the paper, where
by comparison with data on various polymers, we find that
$\alpha \approx a_0 + a_1~M_n + a_2~M_n^2$,
$M_n$ being the average molecular weight of the system.
The constants in this expression are such that $\alpha(M_n)$ decreases
as $M_n$ increases, as discussed above.
Let us now make some more definitions, viz.,
$N$ is the average chain length,  the average chain density is
$\tilde c = \rho N_{avogadro}/(\mu_0 N)$,
$N_e$ is the average chain length between
consecutive points of entanglement, 
the entangled chain number density 
$c_e = \rho N_{avogadro}/(\mu_0 N_e)$,
and the monomer number density $c_0 = \rho N_{avogadro}/\mu_0$,
with $\rho$ being the mass density of polymer, 
$N_{avogadro}$ the Avogadro's number,
and $\mu_0$ the molecular weight of the monomer.
${\l}$ is the length scale in our theory
and we shall take it to be $\l = \lambda \tilde c^{-1/3}$,
where $\lambda$ is a parameter taken to be $(2/3)^{1/3}$ 
as it leads to an expression for the tensile modulus 
in the Gaussian approximation which agrees
with the standard Wall theory result in the limit of short (entangled) states.
We shall use the length $\l$ to scale all other lengths in the system.
These ideas are slightly similar to Stillinger's in another 
context\cite{stillinger,carraro}.
By choosing our length scale in this fashion, it allows us to see how
higher order corrections beyond the Gaussian approximation lead to an
enhanced elastic modulus.
In this manner we have attempted to relate our theory in
an intimate fashion to the notion of entanglement.


\section{Beyond the Gaussian approximation.}

We shall now use diagrammatic methods to go beyond the Gaussian
approximation to the structure function
described at the end of the previous section.
The reason is to be able to describe the crossover
to a state of entanglement.
As discussed in the previous section, the onset of entanglement
is not a phase transition, but simply a crossover.

Figures 1a and 1b show the basic vertices in our theory.
The figure captions describe the Feynman rules which go with
these vertices.
We shall compute only the first non-vanishing terms which arise
from each of these vertices.
The first order contribution of the cubic term is zero, as follows
from symmetry considerations.
We have to go to the second order in the cubic term
to obtain a {\it tadpole} diagram which is non-vanishing,
as shown in Figure 2a.
It serves to renormalize the correlation function in the long
wavelength limit.
Figure 2b is the other {\it setting sun} diagram which comes
from the second order contribution of the cubic term.
It may be expanded in powers of its argument $k$.
The term proportional to $k^2$ helps to renormalize the
{\it diffusion} constant $g$, and serves to diminish it,
as one would expect entanglement to.
Figure 3 shows the conventional {\it bubble} diagram
coming from first order perturbation theory with the
quartic term.
It serves to renormalize the correlation function in the long
wavelength limit.

In order to render the integrals in our theory finite in 
three dimensions,
we shall use the following regularization scheme.
We shall perform an expansion of the denominator of
the Gaussian structure factor in powers of $k$.
We shall retain terms up to ${\cal O}(k^6)$.
This is essentially an expansion in inverse
powers of $\alpha$.
This expansion yields the requisite higher order
terms in the denominators of the Green's function
to render our integrals finite, while ensuring that
$\hat {\cal S}_0(k) > 0$.
This method has the advantage of retaining the correct
long-wavelength behavior, at the expense of high momentum
behavior.
This is acceptable, since we do not expect our theory to be correct
at small wavelengths in any event.
We shall shortly compare our method with the conventional
method of renormalization via counter-terms.
Our single-particle Green's function in the Gaussian 
approximation is now taken to be:

\begin{equation}
\hat {\cal S}_0(k) = {1\over 1 + 2 \alpha' k^2 -2 k^4 + k^6/(2 \alpha')}
\label{Sr}
\end{equation}

With this definition, the net contribution from diagrams
shown in Figures 2a and 3 is:

\begin{equation}
\Sigma_{2a+3}(\alpha) = -\left({3\over 4}\right)~{\cal S}_0(0)
\label{so}
\end{equation}

Figure 2b yields a $k$-dependent contribution to the self 
energy:

\begin{eqnarray}
\Sigma_{2b}(\vec k) &&= \left({1\over4}\right)
        ~\int {d^3 k'\over(2 \pi)^3} \hat {\cal S}_0(k')
         \hat {\cal S}_0(\vert \vec k' - \vec k \vert) \nonumber\\
&&\approx \delta a +\delta g k^2 + {\cal O} (k^4)
\label{skk}
\end{eqnarray}

where:

\begin{eqnarray}
\delta a(\alpha) &&\approx {1\over 128~2^{1/4} \pi \alpha^{3/2}} \nonumber\\
\delta g(\alpha) &&\approx \left({1\over 256 \pi \alpha^{1/2}}\right) 
 \left({{5}\over{2^{3/4} \alpha^2}} - \sqrt 2 \right)\nonumber\\
\label{rr}
\end{eqnarray}

where the integrals were performed by approximating the denominator
of the Gaussian Green's function by terms up to ${\cal O}(k^2)$,
as this suffices to guarantee convergence of the integrals,
so that there is no sensitivity to the higher order terms neglected.
We find that the contribution from Eqn.\ref{so} is extremely small
compared to $\delta a$ from Eqn.\ref{skk}.
This is basically what happens in the usual renormalization scheme,
where one eliminates terms such as ${\cal S}_0(0)$ (when Eqn.\ref{sk}
is used to perform the calculations) using appropriate counter-terms in 
the energy functional.
In fact, in this scheme, $\hat {\cal S}_0(k)$ decays quadratically
with $k$, and the considerations used to obtain Eqn.\ref{rr}
automatically obtain.

With these expressions, we see that the renormalized value
$g_R = g - \delta g$ of the diffusion constant decreases as
$\alpha$ is decreased.
As discussed in section II, the non-local attractive term in our free
energy was identified with the formation of knotty configurations, or
clusters\cite{stepanek}. 
We will therefore identify $g_R$ with the dynamics of such clusters.
Cluster dynamics have been recently observed using state-of-the-art
techniques by Stepanek et al\cite{stepanek}.
$g_R$ is zero near $\alpha = 0.18$.
Note that $\alpha$ decreases 
as we increase $N$ the average chain length.
We thus see that as 
entanglement increases,
the effective diffusion constant decreases, analogous to {\it critical}
slowing down.
We are not aware of explicit experimental observations regarding the
dynamics of clusters, with which $g_R$ has been identified, near the
cross-over threshold.
This effect is similar to the considerations of
Kavallis and Noolandi\cite{kn}, Kroy and Frey\cite{kf},
and Broderix et
al,\cite{broderix}, who study the vulcanization transition in the 
mean field approximation, and find the diffusion
constant going to zero as the vulcanization transition is approached.
They point out that this result agrees with experimental results.
Given the analogy between vulcanization (chemical cross-linking)
and physical entanglement, we believe this result should be experimentally
observable in entangled systems as well.
Broderix et al obtained $g_R \to 0$ linearly with the
average concentration.
We have obtained a more complicated dependence on the concentration.
We find that $\alpha\approx 0.18$ when the renormalized
diffusion constant goes to zero.
One could estimate $\alpha$ using
results from the next section on the tensile modulus of polymers
and experimental
values for polymeric elastic moduli, and then obtaining a value for 
the critical $N^*$ at which the effective diffusion constant goes to zero.

The origin of $\delta g > 0$ can be traced back to 
the nonlocal attractive term in $U_{{\rm eff}}$, defined in Eqn.\ref{gauss}.
This nonlocal attractive term, which we interpreted as
giving rise to entanglement, is responsible for a physical
signature of the onset of entanglement, with $g_R \to 0$.


\section{Tensile modulus.}

It is well-known that the Wall theory result
for the tensile modulus, while yielding
the correct trend, does not agree with experimental data
on moduli by a large factor.
Edwards' application of deGenne's reptation model\cite{doi}
provides an enhancement factor over Kuhn's result,
and shows conceptually how entanglement leads to an increase
in the stiffness of the homopolymer system.
We will show in this section how 
to obtain a similar result in our continuum treatment.
More importantly, we will show we can go further, and describe a
crossover, as the mean chain length between entanglements
is decreased, to a regime where the tensile modulus, instead
of remaining fairly constant, begins to increase extremely
rapidly as a function of decreasing $N_e$.
The reptation model is unable to accomplish this,\cite{doied}
as it assumes that the system is already in the entangled state,
and does not account for inter-chain interactions, beyond
assuming a preformed tube.

The Helmholtz free energy in the Gaussian approximation is given
by:\cite{ramond,binney}

\begin{equation}
{\cal F}_G = -{3\over 2} k T V \tilde c {\hat{\cal S}}_0(k=0) =
-{3\over 2} k T V \tilde c \int d^3x{\cal S}_0(x)
\label{fg1}
\end{equation}

We may represent a strained state of the system by
the transformation
$\vec x \to \vec x' = \vec x + \vec u(\vec x)$
in the above equation.
This is possible because ${\cal S}(\vec x)$ in the above equation represents
the density-density correlation function
$<c(\vec r) c(\vec r - \vec x)>$, so that when the system is
strained, $c(\vec r - \vec x)$
in our theory shifts to $c(\vec r' - \vec x')$,
where $\vec r' = \vec r + \vec u(\vec r)$.
For the case of homogeneous deformation, we shall take
$\vec u(\vec x) = \tensor \epsilon \cdot \vec x$,
where $\epsilon$ is the strain tensor.
The strain is assumed to be volume-preserving, so that $d^3x=d^3x'$.
Our approach is similar to that of Castillo and Goldbart\cite{castillo}.
It is now easy to show, using a Taylor series expansion that
the change in the pressure 
$P =-\left( \partial{\cal F}/\partial V \right)_{T,N}$ is:

\begin{eqnarray}
\Delta P_G &&\approx {3\over 2} k T \tilde c {\hat{\cal S}}_0(k=0) 
      \epsilon_{\alpha \beta} \epsilon_{\gamma \delta}
        {\cal D}_{\alpha \beta \gamma \delta} 
   + {\cal O}(\tensor \epsilon^4) \nonumber\\
{\cal D}_{\alpha \beta \gamma \delta} &&= \left(
       \delta_{\alpha \beta} \delta_{\gamma \delta} +
            \delta_{\alpha \gamma} \delta_{\beta \delta} \right)
\label{sss}
\end{eqnarray}

where the subscript G denotes the Gaussian approximation.
As needed, we can consider the expansion of Free energy to include
higher orders of the strain tensor\cite{bard}.
$\Delta P_G$ is a measure of the change per unit volume of
the energy of the system under strain.
In analogy with a simple harmonic oscillator, the force tensor 
which constrains the system from undergoing a strain $\tensor \epsilon$ 
is given by 
$- 3 k T \tilde c {\hat{\cal S}}_0(k=0) \tensor \epsilon 
          : \tensor {\tensor {\cal D}}$.
Thus, the stress $\tensor \sigma$ required to produce this strain is:

\begin{equation}
\tensor \sigma = 3 k T \tilde c {\hat{\cal S}}_0(k=0) \tensor \epsilon
          : \tensor {\tensor {\cal D}}
\label{stress}
\end{equation}

One may now readily write down the tensile modulus as:

\begin{equation}
Y_G = 3 k T \tilde c 
\label{fg2}
\end{equation}

This is identical to the well-known Wall theory result obtained by in
the limit of  
short (unentangled) chains.
And it's origin is purely entropic.
Our goal is to go beyond this result, and to do that,
we shall {\it dress} the
bare propagator ${\hat{\cal S}}_0(k)$ using the diagrams shown in
Figures 2 and 3.
This immediately leads to the renormalized result $Y_R$:

\begin{eqnarray}
Y_R(\alpha) &&= 3 k T \tilde c {\hat{\cal S}}_R(k=0) 
                     \nonumber\\
      &&= 3 k T \tilde c  \left[
        {1\over{1-\Sigma_{2a+3}(\alpha) - \delta a(\alpha)}}\right]
\label{BR}
\end{eqnarray}

The first of these equations is similar to the connection made between
the structure factor in the long wavelength limit and the bulk modulus
by Kirkwood\cite{kirkwood}.

The result of plotting
the {\it entanglement} factor ${\cal Z} = Y_R/Y_G - 1$
as a function of $\alpha(N)$
is presented in Figure 4.
We see that the enhancement factor,
which is fairly constant above $\alpha=0.2008$,
begins to increase 
dramatically below this value of $\alpha=0.2008$.
As discussed in section II, decreasing $\alpha$ is equivalent to
increasing $N$, the average number of links in a polymer.
And increasing $N$ is associated with increasing entanglement.

To see better the connection between our approach and the underlying
chain parameters, we identify the prefactor $({\cal Z}(\alpha)+1)$
multiplying the Wall theory result with the enhancement obtained within the
reptation model\cite{doied}, i.e., ${\cal Z}(\alpha)+1 \equiv
\left(N b^2/a^2\right)$, 
where $b$ is the monomer length, and $a$ is the diameter of the tube in
the reptation model.
This identification gives us a relation between $\alpha$ 
and the parameters of reptation theory.
It also gives us a relation between our theory and the underlying
chain parameters such as $N$, the average number of links, and $N_e$,
the average 
number of links between successive points of entanglement, viz., 
$N_e^{-1} = \left(b^2/a^2\right)$.
In fact, we can provide such an analytic, approximate relation in the
following manner.
Based on numerical estimates of ${\cal S}_0(0)$, we find that in the
the highly entangled state, 
$\vert\delta a(\alpha)\vert >> \vert \Sigma_{2a+3}(\alpha)\vert$. 
Then, using Eqn.\ref{rr}, we immediately obtain:

\begin{eqnarray}
\alpha(N_e,N) &&\approx \kappa (1-N_e/N)^{-2/3}
\nonumber\\
\kappa &&= \left({1\over
{128~2^{1/4}~\pi}}\right)^{2/3}
\label{scaling}
\end{eqnarray}

It should be noted that this particular scaling relation
Eqn.\ref{scaling} holds only in the 
limit of highly entangled states.
While we have denoted the explicit dependence on $N_e$ and $N$
separately, it should be noted that $N_e$ itself depends on the chain
length $N$.
More generally, the advantage of our theory is that the enhancement
factor $\left[{\cal Z}(\alpha(N))+1\right]$
can change continuously from a value of unity in the unentangled
state, to a fairly large number as the system becomes increasingly
entangled.
As discussed earlier,
the reptation model on the other hand presupposes the formation of
tube-constraint caused by polymer chains surrounding any given chain.
As such it applies only in the highly entangled state.

We introduced our theory in section II in a phenomenological fashion,
and so the best way to
establish the validity of our theory is to 
compare the results of our theory with experiment,
the connection with the underlying chain parameters
(Eqn.\ref{scaling}) notwithstanding.
However, we are not aware of experimental data on the elastic
properties of entangled 
polymers detailing in a precise manner the cross-over to the entangled 
state.

Nevertheless, we were able to find an experimental paper by Onogi et
al\cite{onogi} which provides systematic 
viscoelastic data on amorphous polystyrene in the melt state, over a
wide range of frequencies and molecular weights (and having a low degree
of polydispersity).
We were able to obtain static shear
moduli from the plateau moduli given in this paper.
The conventional assumption of incompressibility then says that
Young's modulus is 
three times the shear modulus.
Armed with these data, we were able to obtain a parameterization of
$\alpha$ as function of the molecular weight $M_n (= \mu_0 N)$ (see
Table 1 and Figure 5).
This parameterization is consistent with the stress-strain data
provided by Bicerano et al\cite{bicerano}.
This parameterization is also consistent with Eqn.\ref{scaling}.
Equation \ref{scaling} implies that for very large molecular weights,
$\alpha \to \kappa \approx 0.01635$.
From Figure 5,
we see such
values for $\alpha$ occurring
at the high 
end of the range of molecular weights.
The parameterization depicted in Figure 5 may be thought of as a power 
series expansion of Eqn. \ref{scaling}.

The cross-over between the un-entangled state and entangled state
appears to take place between $M_n \approx 113,000$ and 
$M_n \approx 60,000$, as the plateau in the storage moduli
measurements disappears somewhere 
between these two values of the molecular weights\cite{onogi}.
Unfortunately, this cross-over is not
described well by these data, as measurements in the cross-over region are
unavailable.
This could well be due to the critical slowing down which our theory
predicts will occur during the cross-over (see section III).
As can be seen from Table 1 and Figure 5, the data develops noise
at the low end of the range of molecular weights.
And this is consistent with the prediction of critical slowing
down, as the system takes very long to settle into a metastable state.
Since we have a cross-over phenomenon, it is difficult to pinpoint a
single value of $M_n$ at which the transition to the entangled state
occurs. 
But based on Figure 6, in which the logarithm of the enhancement factor
has been plotted as 
a function of $M_n$, we see that it starts to become appreciable ($>>
1$) in the neighborhood of $M_n \sim 50,000$.
The theory thus predicts that it is in this neighborhood that the
cross-over takes place.
It would be very interesting to see more experiments performed in the
future, say for amorphous polystyrene, with $M_n$ in the range just
discussed, to see if the path of the cross-over agrees with 
that shown in Figure 6.
It is even more important to do such experiments to ascertain 
if the critical slowing down predicted in section III does indeed occur.


\section { Conclusions}

We postulated an extension of the Cahn-Hilliard functional to describe
entanglement in polymers.
We extended the Cahn-Hilliard
functional with two terms.
One is an attractive nonlocal term which describes the effect of
entanglement,
and the other a local repulsive term indicative of excluded volume 
interactions.
We showed in Appendix A how this extended functional can be derived
using notions from gauge theory.
Using field theoretic techniques to go beyond the Gaussian approximation,
we showed that the onset of entanglement is a crossover phenomenon,
signaled by the effective diffusion constant going to zero.
We developed a simple model to connect the single parameter in our theory
with the parameters of the underlying chains by comparison with
available experimental data on amorphous polystyrene melts.
A reasonable estimate for the critical partial chain concentration
at which this crossover occurs was obtained.
While this is consistent with available data, further experiments to
determine the details of this cross-over were suggested, especially to 
ascertain if the critical slowing down predicted in this paper does
indeed occur.


\section {Acknowledgments}

I would like to acknowledge discussions with Brian Kendrick
regarding the gauge theoretic formulation of the problem.
I would also like to thank Dr. J. Bicerano for kindly supplying
raw experimental stress-strain data\cite{bicerano} on amorphous
polystyrene utilized 
for a consistency check of the theory developed in this paper.
The work described in this paper was done under aegis of the LDRD-CD
program on polymer aging at the Los Alamos National Laboratory.


\appendix
\section{}

In this appendix, we shall show how to derive Eqn.\ref{gauss},
along with the choice of parameters made in section II,
using notions from gauge theory.
Let us start with:

\begin{equation}
\beta u_0 ({c}({\bf s}))  =  \left({g\over2}\right) 
      {{\partial {c}({\bf s})}\over{\partial s_i}}
                          {{\partial {c}({\bf s})}\over{\partial s_i}}
\label{0}
\end{equation}

where the variables have been defined in section II.
We will use this form as our starting point to generate a more
complete energy functional using gauge invariance.

From Eqn.\ref{0} we see that $u_0$ is invariant under
global translations of $c(\vec s)$, i.e. under
$c(\vec s) \to c(\vec s) + h$ where $h$ is a constant.
And the appropriate group to consider is T$_1$.
The physical origin of this group can be traced back to the fact that the
quadratic (positive, semi-definite) form of the energy density 
is dictated by expanding the internal energy around a minimum, in a 
Landau-like fashion.
Physicality of
T$_1$ transformations demands that 
$c' + c_e >$ 0,
where $c'$ denotes the deviation of the specie concentration
from its average, and that number conservation is guaranteed.
These physical constraints will be incorporated into the evaluation
of the partition function, {\it after} gauging $U_0$,
in a manner similar to applying gauge constraints in QFT.

Our physical motivation for seeking local gauge invariance 
under
T$_1$ is the same as that of Yang and Mills\cite{YM}, and 
in quantum electrodynamics (QED), where 
one observes the invariance of the noninteracting Lagrangian
under certain global transformations.
One then demands covariance of the theory when these symmetry
operations are {\it local} i.e., when the transformations are
space-time dependent.
A reason for this, as given by Yang and Mills, is that one can 
now freely interchange between the fields as one moves through
space and time, while leaving the physics covariant.
It is important to note that gauge theory in QFT is {\it not} a
result of the fact that the phase of the field is
not measurable. 
In fact, Aharonov and Bohm showed many years ago that
the phase in quantum mechanics is indeed observable.
Local transformations under T$_1$
generate concentration fluctuations
which arise from entanglement.

Following Yang and Mills,\cite{YM} local gauge invariance of $u_0$
under T$_1$ motivates us to define new fields $\bf b$, which have 
invariance properties appropriate to T$_1$.  
We define a covariant derivative 
${\partial\over{\partial s_i}} \to ({\partial\over{\partial s_i}} +
                                         q \tau b_i)$,
where $\tau = {\partial\over\partial c}$ is the generator of T$_1$, 
$q$ is a `charge', or
equivalently, a coupling constant, and the $b$-fields are analogs
of the magnetic vector potential in electrodynamics. 
In our previous theory of self-assembly\cite{shirish},
gauge fields arose from
the underlying covalent bonds between the two species in the system.
In the present case, where we wish to describe entanglement, the 
gauge fields are to be thought of as arising purely from
statistical considerations alone.
On the other hand, chemical cross-linking would provide a 
physical origin for the $b$-fields in vulcanized systems.
The energy functional
for the $b$-fields is defined 
$\grave {\rm a}$ la Yang and Mills, via the minimal
prescription.  
With this, our original internal energy density is transformed
into:

\begin{equation}
\beta u_0 \to \beta u = \beta u_0 + \beta u_{int} + \beta u_{YM} 
\label{}
\end{equation}

where $u_{int}$ refers to the interaction energy density, and $u_{YM}$
is the energy density associated with the Yang-Mills $b$-fields alone.
Equivalently, we may define the total energy functionals associated
with these energy densities:

$$
\beta U_0 \to \beta U = \beta U_0 + \beta U_{int} + \beta U_{YM},
$$

where

\begin{equation}
\beta u_{int} = \vec J ({c}) \cdot \vec b ({\bf s}) +
           f~ \vec b ({\bf s}) \cdot \vec b ({\bf s}) 
\label{}
\end{equation}

with

\begin{equation}
\vec J ({c}) = g q \vec \nabla c
\label{}
\end{equation}

\begin{equation}
f = \left({g q^2\over2}\right) 
\label{}
\end{equation}

We need one more definition for completeness:

\begin{equation}
\beta u_{YM} = \left({1\over{4}}\right)~
\left( {\partial b_i\over \partial s_j}-{\partial b_j\over \partial s_i}\right)
\left( {\partial b_i\over \partial s_j}-{\partial b_j\over \partial s_i}\right)
\equiv {\vec {\cal B}^2\over 4}
\label{}
\end{equation}

This equation can be cast into the following form:

\begin{equation}
\beta u_{YM} = -\left({1\over{2}}\right)~b_i {\nabla}^2 b_i
\label{ibp}
\end{equation}

Eqn.\ref{ibp} is obtained via an integration by parts, in the transverse gauge.
Since we are dealing with an Abelian gauge theory, it is permissible to
insert this transverse gauge manually, without resorting to the formal
machinery of Faddeev and Popov.

Again using the transverse gauge and integrating by parts, 
it is clear that:

\begin{eqnarray}
\int d^3 s \vec \nabla c(\vec s) \cdot \vec b(\vec s) 
  &&= -\int d^3 s~c(\vec s)~[\vec \nabla \cdot b(\vec s)] \equiv 0 \nonumber\\
&&=i~\int d^3 s \vec \nabla c(\vec s) \cdot \vec b(\vec s)
\label{cruc}
\end{eqnarray}

It is this crucial identity which allows us to get the
precise form for Eqn.\ref{gauss}, which we motivated
in section II using an intuitive approach.
It is the nature of the T$_1$ group which permits this manipulation
to go through successfully.
In our earlier paper\cite{shirish},
where we used the SO(2) group, such a manipulation would not have 
availed us any advantage.

Note that we are utilizing a non-relativistic version of the
Yang-Mills procedure, since we are only concerned with time-independent
problems.
Furthermore, since we are concerned with translations in T$_1$,
there is only a single generator to contend with,
so that the resulting functional is only quadratic and not quartic
in the {\it b}-fields.

It is important to emphasize that the usual application of the Yang-Mills
procedure in QFT implies the existence of fundamental
interactions.  In our case, we are applying the principle of local
gauge invariance at the $mesoscale$.  Consequently, we do not expect to
discover any new fundamental interactions by using gauge invariance.
Rather, we interpret the new $b$-fields as yielding correlations between the
concentration fields.  
These correlations could also be thought of as effective interactions,
which arise at the mesoscale from the underlying electrostatic
interactions between molecules.

The partition function we need to evaluate is now:

\begin{equation}
Q' = \int {{\cal D}c} 
            ~\theta\left(c\right) 
                      \prod_{k=1,3}{{\cal D}b_k}
                               ~exp-\beta (U_0+U_{int}+U_{YM})
\label{Q}
\end{equation}

Equation \ref{Q} is a functional integral,
where the step functions denoted by $\theta$ imply that we must 
restrict integration to positive semi-definite values of
the fields.

Since the $b$-fields appear only quadratically in the above
functional, it is straightforward to integrate over them, and obtain
an effective internal energy functional involving only ${c}$,
upon using Eqn.\ref{cruc}.
The result is:

\begin{eqnarray}
&&\beta U_{{\rm eff}} = \beta U_0 + \beta \Delta U_{{\rm eff}} \nonumber\\
        &&= \beta U_0 +{1\over{4}} \int {\rm d}^3s \int {\rm d}^3 s'~
                        J_i({c}({\bf s}))~
         \left(
            {1\over{f -
                              {1\over{2}}\nabla^2
                   }
            }\right)_{{\bf s},{\bf s}'}~
                                   J_i({c}({\bf s}'))
\label{}
\end{eqnarray}

Note that in doing so, we have ignored an overall trivial normalization 
constant that appears in the evaluation of the partition function $Q'$.
This is permissible, as this factor cancels during the evaluation of 
averages of observable quantities.

To reveal the physics in this effective functional, we perform
some straightforward algebra to write our result as:

\begin{eqnarray}
&&\beta U_{{\rm eff}} = \beta U_0
  +\left({\alpha^2\over 2}\right) \int {\rm d}^3 s~{{c}}({\bf s})
                                       {{c}}({\bf s}) \nonumber\\
  &&- \left({\alpha^4 \over {2 \pi}}\right) \int {\rm d}^3 s \int {\rm d}^3 s'~
                     {{c}}({\bf s}) {exp(-\sqrt{2 \alpha^2/g}
                                        \vert {\bf s} - {\bf s}' \vert)\over
                                              {\vert {\bf s} - {\bf s}' \vert}
                                           }
                                                 ~   {{c}}({\bf s}')
\label{gauss1}
\end{eqnarray}

where $\alpha^2 = g^2 q^2 /2 $ (we shall use units in which q=1).
Note that $U_{{\rm eff}}$ is quadratic, the generator of T$_1$ making
sure that higher order terms do not appear in our functional.
Equation \ref{gauss1} is one of the main results of our paper,
and provides a deeper motivation for the model developed in
section II on intuitive grounds.
As discussed in section II, the form 
of Equation \ref{gauss1} guarantees number conservation.
Equation \ref{gauss1} shows that entanglement may be understood
in the context of a mesoscopic gauge theory.
Note that the two terms we just discussed have signs opposite
those of corresponding terms in theories of self-assembly\cite{shirish}.
We thus see that using T$_1$ instead of SO(2) in the previous
theory\cite{shirish} has led to a qualitatively different
theory.
Finally, this approach provides an alternative gauge
theory of polymer entanglement than the one given 
by Brereton\cite{brereton}.
In this paper, Brereton works explicitly with entangled strands of
polymers and provides a static theory which yields e.g. a renormalized 
expression for the radius of gyration.
It would be interesting to see how Brereton's theory could be
generalized to include dynamics as well.



\newpage
\mediumtext
\begin{table}
\caption{Fitted values of $\alpha$ obtained in comparison with
  experimental data from Onogi et al on amorphous polystyrene melts at 
  160 $^{\circ}$C.  Young's modulus is denoted by $Y$.}
\begin{tabular}{ccc}
\text{Molecular weight}&\text{Y$_{exptl}$  (psi)}&\text{$\alpha$(fitted)}\\
\tableline
581,000 & 93 & 0.0166 \\
513,000 & 84  & 0.0167\\
351,000 & 66 & 0.0171\\
275,000 & 65 & 0.0173 \\
215,000 & 62 & 0.0177\\
167,000 & 56 & 0.0184\\
113,000 & 106 & 0.0179\\
\end{tabular}
\label{table1}
\end{table}

\begin{figure}
\caption{(a) is a pictorial representation of the cubic term in $A$.
Each leg corresponds to a factor of $c$, the field.
The intersection of the three legs symbolizes a factor of $\gamma=1/6$, 
the coupling constant.
(b) is a pictorial representation of the quartic term in $A$.
A factor of $-1/12$ is to be inserted at the intersection.}
\label{fig1}
\end{figure}
\begin{figure}
\caption{ (a) represents the {\it tadpole} diagram which is crucial in our
calculations.
(b) represents the {\it setting sun} diagram.
Both (a) and (b) are second order contributions to the correlation
function coming from the cubic interaction term, the first order
corrections being null.}
\label{fig2}
\end{figure}
\begin{figure}
\caption{ This figure represents 1-loop (bubble) contribution from the quartic 
interaction term in $A$.}
\label{fig3}
\end{figure}
\begin{figure}
\caption{ This is a plot of ${\cal Z}=Y_R/Y_G - 1$ as a function of $\alpha$.
Notice that the factor is virtually constant above $\alpha = 0.20$,
followed by a dramatic
increase below this value of $\alpha$.
Approximations employed in the calculation cause the entanglement factor
to diverge at $\alpha \approx 0.01$.
Remember, decreasing $\alpha$ corresponds to increasing entanglement.
}
\label{fig4}
\end{figure}
\begin{figure}
\caption{This dots are the values of $\alpha$ required to
  achieve agreement with experimental values of moduli of amorphous
  polystyrene melts for a range of $M_n$.  The solid line indicates a
  least squares fit obtained using $\alpha (M_n) = a_0 + a_1 M_n + a_2 
  M_n^2$, with $a_0=0.0189,~a_1=-6.18\times 10^{-9},~a_2=3.89\times
  10^{-15}$.
Caution should be exercised in using this expression to estimate
$\alpha$ too far from the range in which the fit was made. }
\label{fig5} 
\end{figure}
\begin{figure}
\caption{Plot of Log$\left[{\cal Z}\right]$ as a function of the
  molecular weight $M_n$.  Notice that ${\cal Z}$ begins to attain
  nontrivial values above $M_n \sim 5\times 10^4$.  The curve indicates
  the path of the cross-over from an un-entangled state to an entangled state.}
\label{fig6}
\end{figure}
\newpage


\input epsf.tex
\leavevmode

\centerline{Chitanvis,\ \ Fig.\ \ref{fig1}}
\vspace{0.1cm}
\hspace{0.1cm}\epsfbox{ent.eps.1}
\eject

\centerline{Chitanvis,\ \ Fig.\ \ref{fig2}}
\vspace{0.1cm}
\hspace{0.1cm}\epsfbox{ent.eps.2}
\vfill\eject

\centerline{Chitanvis,\ \ Fig.\ \ref{fig3}}
\vspace{3.0cm}
\hspace{0.1cm}\epsfbox{ent.eps.3}
\vfill\eject

\centerline{Chitanvis,\ \ Fig.\ \ref{fig4}}
\vspace{0.1cm}
\hspace{0.0cm}\epsfbox{ent.eps.4}
\vfill\eject

\centerline{Chitanvis,\ \ Fig.\ \ref{fig5}}
\vspace{0.1cm}
\hspace{0.0cm}\epsfbox{ent.eps.5}
\vfill\eject

\centerline{Chitanvis,\ \ Fig.\ \ref{fig6}}
\vspace{0.10cm}
\hspace{0.0cm}\epsfbox{ent.eps.6}
\vfill\eject


\end {multicols}
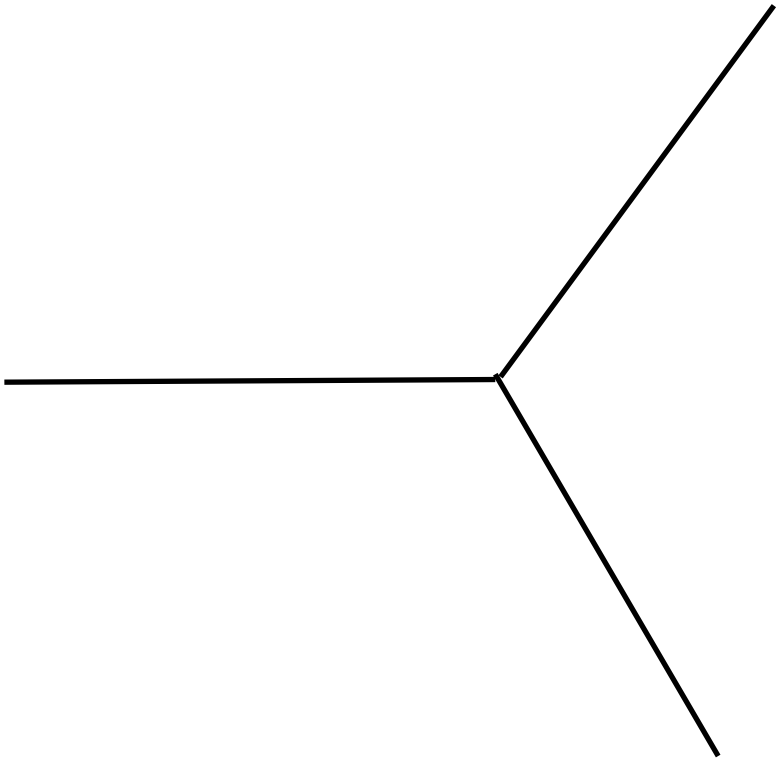
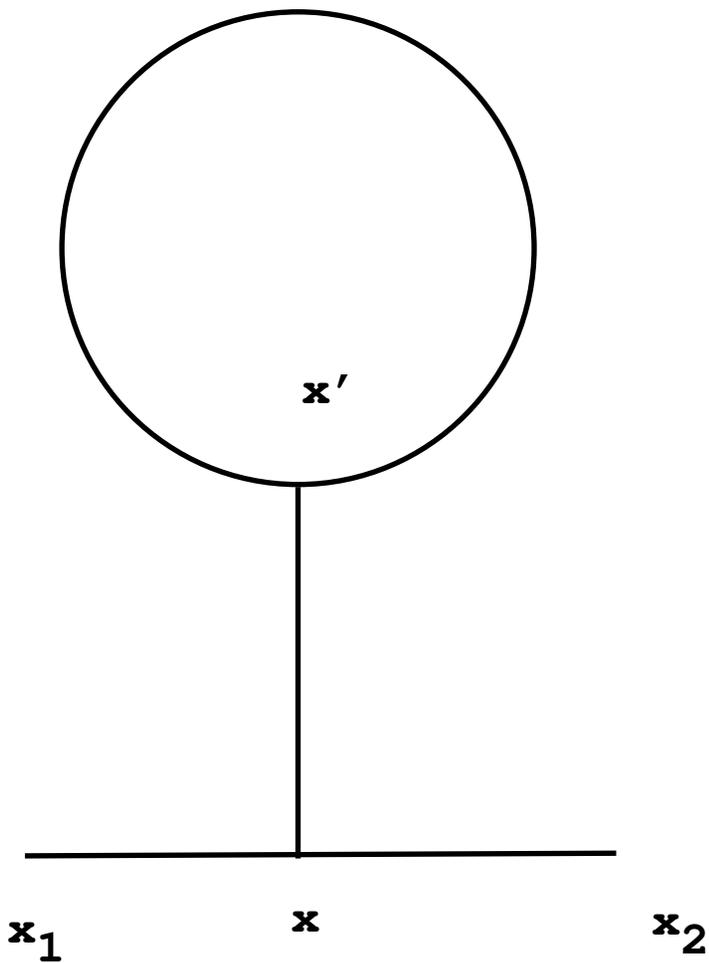
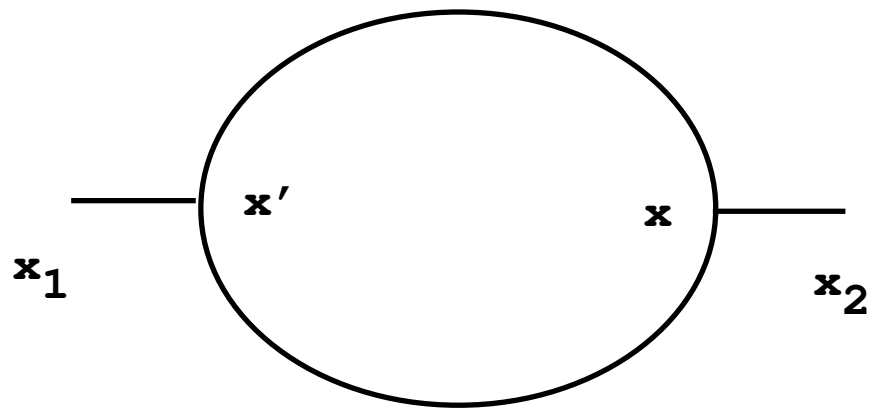
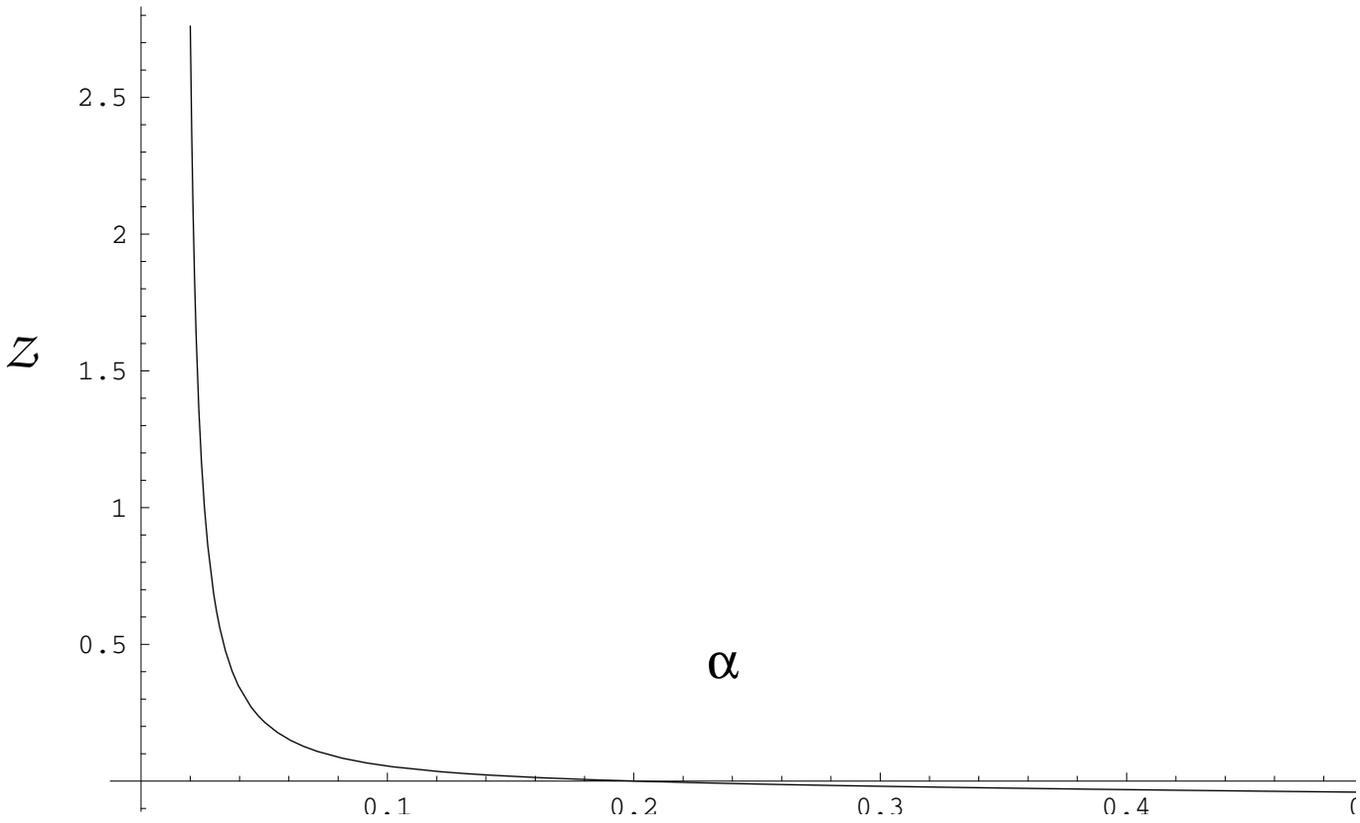
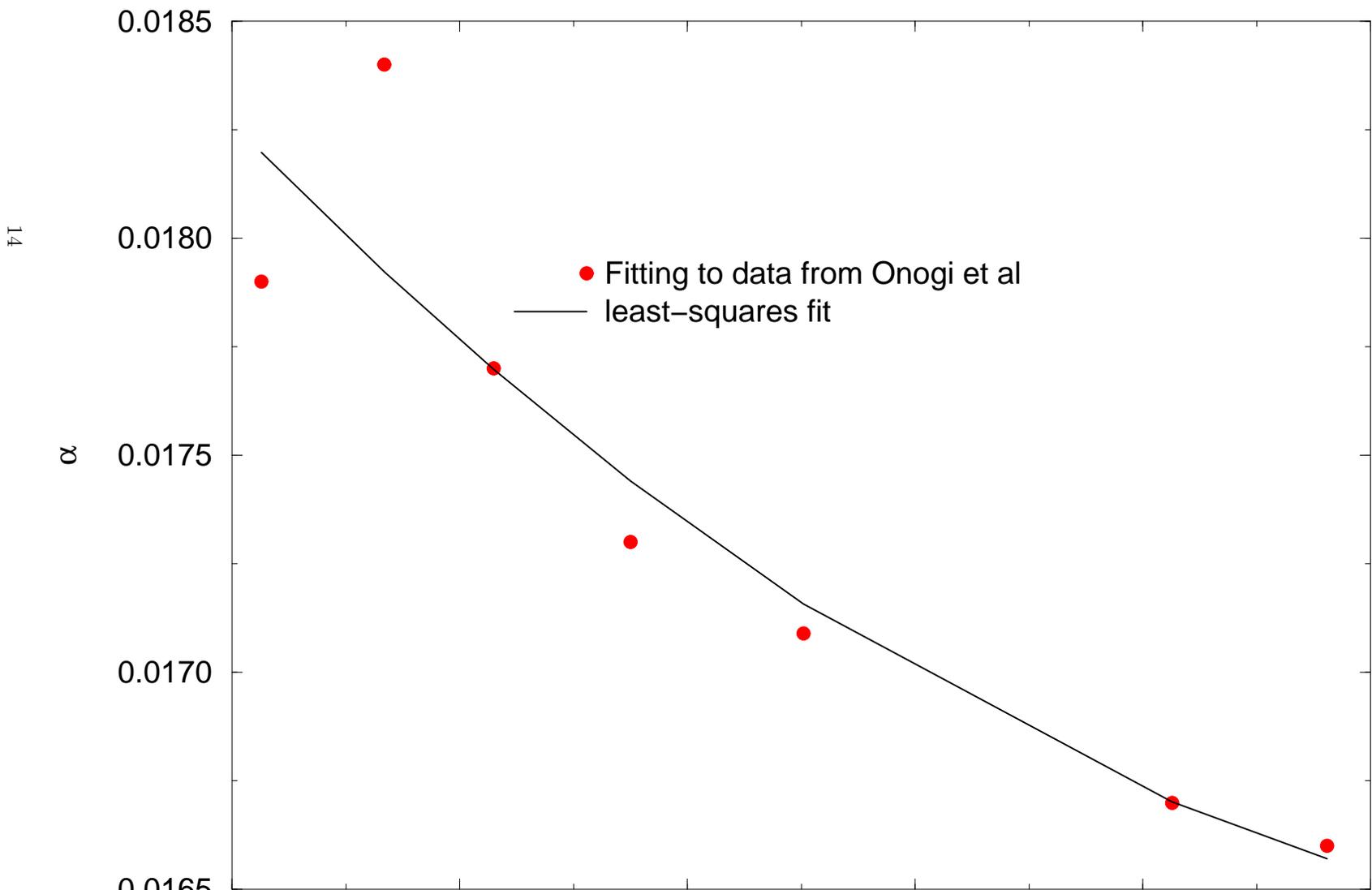
\end{document}